\documentstyle[aps,multicol,subeqn,epsf,rotate,citesort]{revtex}
\newcommand{\gl}[1]{Eq. (\ref{#1})} 
\newcommand{\gls}[2]{Eqs. (\ref{#1},\ref{#2})} 
\newcommand{\glto}[2]{Eqs. (\ref{#1}) to (\ref{#2})} 

\def\gtrless{\raise2.5pt\hbox{$>$}\llap{\lower2.5pt\hbox{$<$}}}
\def\gtrapprox{\raise2.5pt\hbox{$>$}\llap{\lower2.5pt\hbox{$\approx$}}}
\newcommand{\bsq}[1]{\begin{subequations}\label{#1}}
\newcommand{\esq}{\end{subequations}}
\newcommand{\beq}[1]{\begin{equation}\label{#1}}
\newcommand{\eeq}{\end{equation}}
\newcommand{\beqa}[1]{\begin{eqnarray}\label{#1}}
\newcommand{\eeqa}{\end{eqnarray}}
\newcommand{\fur}{\qquad\mbox{for }\, }

\newcommand{\om}{$\omega_q$\ }
\newcommand{\sq}{$S_q$\ }
\newcommand{\hq}{$h_q$\ }
\newcommand{\gr}{$g(r)$\ }
\begin{document}

\title{Inter--molecular structure factors of macromolecules in solution:
integral equation results}
\author{M.~Fuchs$^{(a)}$ and M.~M\"uller$^{(b)}$}
\address{$^{(a)}$ Physik-Department, Technische Universit{\"a}t M{\"u}nchen,
85747 Garching, Germany;\\
$^{(b)}$ Institut f\"ur Physik, Johannes Gutenberg--Universit\"at Mainz,
55099 Mainz, Germany }
\date{April 19, 1999}
\maketitle

\begin{abstract}
%\noindent
The inter--molecular structure of semidilute polymer solutions is studied
theoretically. 
The low density limit of a generalized Ornstein--Zernicke integral equation
approach to polymeric liquids is considered. Scaling laws for the
dilute--to--semidilute crossover of random phase (RPA) like structure
are derived for the inter--molecular structure factor on large
distances when inter--molecular excluded volume is incorporated at the
microscopic level. This leads to a non--linear equation for the excluded volume
interaction parameter. For macromolecular size--mass
scaling exponents, $\nu$, above a spatial--dimension dependent value,
$\nu_c=2/d$, mean field like density scaling is recovered, but for $\nu<\nu_c$
the density scaling becomes non--trivial in agreement with  field theoretic
results and justifying  phenomenological extensions of RPA. The structure of
the polymer mesh in semidilute solutions is discussed in detail and
comparisons with large scale Monte Carlo simulations are added.
Finally a new possibility to determine the correction to scaling exponent
$\omega_{12}$ is suggested.
\bigskip

{\noindent PACS numbers: 61.25.Hq, 61.12.Ex, 61.20.Ja }
\medskip

\end{abstract}

\begin{multicols}{2}
 
\section{Introduction \label{sex1}}

Whereas the
conformational statistics of a single flexible polymer chain in dilute and
semidilute solutions are understood rather well, less is
known about the inter--molecular packing.  It is well understood that a
semidilute polymeric solution builds up a temporary mesh with a
mesh size, the density screening length $\xi_\rho$, which
for macromolecular solutions can become large compared to the length scales
characterizing the individual monomers \cite{degennes,descloi}. 
However, the inter--molecular packing inside the mesh but still on length
scales 
large compared to the chemistry dependent local length scales is as  yet
unclear. It has been the focus of recent neutron scattering experiments
\cite{ullman,jannink}, scaling considerations and field theoretic calculations
\cite{jannink,schaefer}, and of computer simulations \cite{yamakov}. Older
theories for the inter--molecular structure, which either used the random phase
approximation (RPA) \cite{benoit}
 or the assumption of Gaussian inter--molecular correlations
\cite{bueche}, failed to incorporate the non--mean field like correlations on
scale $\xi_\rho$ of semidilute polymer solutions. The recent field theoretic
results lead to contradicting results as will be pointed out and resolved in
this contribution.  

Integral equation theories for simple liquids directly address the problem of
inter-particle packing in dense fluids. 
Starting with the work of Schweizer and Curro \cite{kcur1}, this approach has
successfully been extended to macromolecular liquids. The polymer
reference interaction site model (PRISM) integral equations have been
fruitfully applied to describe inter alia the inter--molecule correlations in
dense homopolymer systems, polymer blends and block copolymer melts
\cite{kcur3}. PRISM is a macromolecular generalization of the RISM theory of
small molecules of Chandler and Andersen \cite{chandler,chandl}.
The low density limit of PRISM theory shall be worked out in detail in
this contribution in order to discuss the density correlations on the mesh size
length scale. This works either extends  \cite{thread0,thread,fractal}, or
complements  \cite{threada,aviksem} previous studies. 

It is a priori not related to nor required for the success of the PRISM
approach to polymer melts whether it also correctly
captures the long ranged correlations of semidilute polymer
solutions. As liquid correlations in melts generally are short
ranged, an approach like PRISM appropriate for dense systems, need not be a
useful approach to \mbox{(semi--)}
 dilute solutions, where long ranged correlations
are of interest. Nevertheless, the simplification of the PRISM equations to low
polymer densities worked out here will be argued to provide a useful
description of the inter--molecular correlations building up the polymer mesh
in  polymer solutions \cite{kcur3,thread0,thread}.
 Criteria for the quality of the approach will be
established from comparisons with simulations, field theory
 and mean field results.

The aspect of screening of the intra--molecular excluded volume shall be
neglected in this work. It would require the use of self-consistent PRISM
theory, which is considerably more demanding \cite{kcur3}. Moreover, the errors
made, when neglecting the crossover to Gaussian intra--molecular correlations
on  length scales large compared to the density screening length $\xi_\rho$,
will not affect the scalings of the inter--molecular correlations for
distances smaller than $\xi_\rho$, which are the main focus of this
contribution. Thus, in the following the intra--molecular correlations shall be
characterized by a density independent polymer structure factor \om, which, for
macromolecules of $N$ segments at the positions ${\bf r}_\alpha$,  is defined
as follows: 
\beq{1.0}
\omega_q = 
\frac{1}{N} \sum_{\alpha,\beta =1}^N \langle
e^{i  {\bf q} ( {\bf r}_\alpha  -{\bf r}_\beta ) } \rangle \; .
\eeq
Its full functional form will
not be required. Knowledge of its variation for small, large and intermediate
wave vectors suffices \cite{degennes,descloi}.

For small wave vectors, the number of scattering units, the index of
polymerization $N$, where $N\gg 1$ for macromolecules,
and the global molecular size, the radius of gyration
$R_g$, can be obtained from a scattering experiment measuring \om:
\beq{1.0a}
\omega_q \to N\, ( 1 - \frac 1d \; q^2 R_g^2 + \ldots \, )\;, \fur q R_g \ll
1\; , 
\eeq
where $d$ is the spatial dimension. In an intermediate wave vector range, the
macromolecule is supposed to be self similar, which leads to a power law
behavior in \om determined by the fractal dimension, $d_F=1/\nu$:
\beq{1.0b}
\omega_q \to \frac{1}{(q\sigma)^{1/\nu}} \; , \fur\; 
 1/R_g \ll q \ll 1/\sigma\; .
\eeq
The fractal exponent $\nu$ also determines the size--mass scaling, $R_g \propto
\sigma N^\nu$, where a smooth crossover from \gl{1.0a} to \gl{1.0b} is assumed
around $qR_g\approx 1$. The assumption of an intermediate self--similar
molecular structure rules out the study of compact macromolecules, e. g. hard
sphere like colloids, but is appropriate for polymer chains in good,
$\nu=0.588\ldots$, or $\Theta$--solvents, $\nu=\frac 12$, or for rods, $\nu=1$,
which 
share some properties with semi-flexible polymer molecules like actin or DNA
\cite{descloi,degennes,sackmann}. 
The Kuhn'sche--segment size $\sigma$ in \gl{1.0b} is of
the order of local polymer--specific length scales where microscopic segmental
packing effects influence the complicated structure of \om. 
This chemistry dependent variation of \om around $q\sigma\approx1$ can be
included in PRISM studies \cite{kcur3}, but shall be neglected here. Only the
self scattering contribution, $\alpha=\beta$ in \gl{1.0}, which is the only
remaining contribution for large  wave vectors, $q\sigma \gg 1$, is universal
and needs to be considered.  
\beq{1.0c}
\omega_q \to 1 \; , \fur q \sigma \gg 1\; .
\eeq
Thus, a generic smooth crossover from the point particle self scattering
term, \gl{1.0c}, to the self similar intra--molecular correlations, \gl{1.0b},
will be assumed. Chemistry dependent local packing will show up in all
correlation functions on microscopic length scales but will not, except for
in prefactors,  affect the inter--molecular structure on global length scales
like the molecule size, $R_g$, or the mesh width, $\xi_\rho$; as will be shown
explicitly. 

In order to characterize the total, including the inter--molecular, density
correlations of an interacting polymer system, further correlation functions
need to be introduced. In order to compare them with results from other
approaches it is useful to recall their definition as used in  PRISM theory
\cite{hansen,kcur3}. To be specific, let us consider $n$ polymers with $N$
scattering units in a $d$--dimensional volume $V$, where in the thermodynamic
limit  the number density of segments, $\varrho = \frac{nN}{V}$, is kept fixed;
i. e. $n, V \to\infty$ with $\varrho=$ const.. 
The local, fluctuating density is
\beq{1.1}
\varrho( {\bf r}, t ) = \sum_{i=1}^n \sum_{\alpha=1}^N \delta({\bf r} - {\bf
r}^{(i)}_\alpha(t) )\; ,
\eeq
with equilibrium average $\langle \varrho( {\bf r} , t ) \rangle = \varrho$.
The spatial components of the density fluctuations shall be denoted by
$\varrho_{{\bf q}}$, where:
\beq{1.2}
\varrho_{{\bf q}} = \int d^d r\; e^{i {\bf q} {\bf r}}\; \varrho({\bf r}) =
\sum_{i=1}^n \sum_{\alpha=1}^N e^{i  {\bf q} {\bf r}^{(i)}_\alpha} \; .
\eeq
Their statistical average vanishes except for zero wave vector,
$\langle \varrho_{{\bf q}} \rangle = \varrho \delta_{{\bf q},{\bf 0}}$.

The total structure factor, \sq, as measured for example by coherent neutron
scattering \cite{descloi,hansen}
 is given by the second moment of the wave vector
dependent density fluctuations, from \gl{1.2}:
\beqa{1.3}
S_q & = & \frac{1}{nN} \langle \varrho^*_q \varrho_q \rangle - nN \delta_{{\bf
q},{\bf 0}} \nonumber \\ &  = &
 \frac{1}{nN} \sum_{i,j =1}^n \sum_{\alpha,\beta =1}^N \langle
e^{i  {\bf q} ( {\bf r}^{(i)}_\alpha  -{\bf r}^{(j)}_\beta ) } \rangle - nN
\delta_{{\bf q},{\bf 0}}     \; .
\eeqa
The total density fluctuations are straightforwardly separated into density
fluctuations on the identical polymer, \om, the intra--molecular structure
factor, and on different polymers, \hq, the inter--molecular structure factor.
\beq{1.4} 
S_q = \omega_q + \varrho \,  h_q \; ,
\eeq
where the intra--molecular part has already been defined in \gl{1.0} above.
The inter--molecular structure factor, \hq, describes the packing of different
molecules and is given by the restricted sum $i\ne j$:
\beq{1.5}
h_q = \frac{V}{n^2N^2}  \sum_{i,j =1, i\ne j}^n 
\sum_{\alpha,\beta =1}^N \langle
e^{i  {\bf q} ( {\bf r}^{(i)}_\alpha  -{\bf r}^{(j)}_\beta ) } \rangle
- V \delta_{{\bf q},{\bf 0}}     \; .
\eeq
The inter--molecular structure factor is the Fourier transform of the
inter--molecular pair correlation function, \gr:
\beq{1.6}
h_q = \int d^dr e^{i{\bf q} {\bf r}} \; (\, g(r) - 1 \,) \; .
\eeq
The pair correlation function describes the probability averaged over all
segments of
finding at a distance $r$ from site $\alpha$ on molecule $i$ another segment
$\beta$ of a different molecule $j$. From \gls{1.5}{1.6}, one obtains:
\beq{1.7}
g(r) =  \frac{V}{n^2N^2} \sum_{i,j=1;i\ne j} \sum_{\alpha,\beta =1}^N \langle
\delta( {\bf r} - ({\bf r}^{(i)}_\alpha  - {\bf r}^{(j)}_\beta)  ) \rangle \; .
\eeq
\gr is non--negative and approaches unity for large separations $r$,
because then statistical correlations between the sites at  ${\bf
r}^{(i)}_\alpha$ and  ${\bf r}^{(j)}_\beta$ have vanished \cite{hansen}.

Implicit in the Eqs. (\ref{1.3}) to (\ref{1.7}) is a neglect of a special site
dependence of the density fluctuations as it might for example arise from
chain--end effects for linear polymers \cite{kcur1}. Star polymers, where
sites in the core region possibly experience very different local density  
fluctuations than sites in the star--arms, also would require a more elaborate
treatment \cite{kcur3}. 
Nevertheless, for arbitrary macromolecular architectures, 
the above defined correlation functions
are experimentally measurable at least in principle and can also be determined
from computer simulations. They contain information about the local liquid
structure and remain meaningful in the whole accessible density range, from
dilute solutions to melts.

\section{PRISM integral equations  \label{sex2}}

Whereas the effects of the excluded volume on the intra--molecular structure,
like swelling, are already taken into account in \gl{1.0}, PRISM theory
\cite{kcur1,kcur3} explicitly enforces inter--molecular excluded volume by
requiring the pair correlation function to vanish for distances smaller than
the segment size R. 
\beq{2.1}
g(r) = 0 \fur r < R\; ,
\eeq
Building upon the Ornstein--Zernicke approach so successful for simple liquids
\cite{hansen}, an averaged molecular site--site Ornstein--Zernicke like
equation \cite{chandler,chandl} is formulated
and can be viewed as definition of an effective potential, the direct
correlation function $c_q$: 
\beq{2.2}
h_q = \omega_q c_q \, ( \omega_q +  \varrho h_q ) \; . 
\eeq
Equations (\ref{1.4}) and (\ref{2.2}) can also be brought into a RPA--like form
which supports the interpretation of the direct correlation function as an
effective potential. 
\beq{2.3}
S^{-1}_q = \omega^{-1}_q - \, \varrho c_q \; . 
\eeq
Different from the RPA approach, $c_q$ is not considered to be given, but needs
to be found from a solution of the non--linear integral equations. Besides
Eqs. (\ref{2.1}) to (\ref{2.3}), a further equation, the  ``closure''
approximation, is required to determine the solution.
The most simple and yet appropriate closure to treat
the inter--molecular steric repulsion is the Percus--Yevick (PY) approximation
which  expresses the expectation that the effective potential is short ranged:
\beq{2.4} 
c(r) = 0 \fur r > R\; ,
\eeq
Note, that the interaction described by $c(r)$ is localized on microscopic
length scales. Thus, the PRISM equations describe the interplay of local
inter--molecular steric interactions and long ranged intra--molecular
correlations due to macromolecular connectivity \cite{kcur1,kcur3}.

\subsection{Thread limit for (semi--) dilute solutions \label{sex2a}}

In the simplified non--self-consistent PRISM approach, the intra--molecular
structure \om, from \gl{1.0}, is assumed to be given, and (mostly
numerical) techniques to solve the integral equations,
\glto{2.1}{2.4} are employed \cite{kcur3}. Note that
for a Pade approximation to \om for  Gaussian chains with $\nu=\frac 12$ in
$d=3$ an analytic solution of the PRISM equations on all length scales exists
\cite{threada}. The solution technique employing the Wiener--Hopf factorization
as pioneered by Baxter \cite{baxter} can straightforwardly be extended to
Gaussian chains in  (low) odd dimensions, $d=5, 7, \ldots$, but a simpler
approach can also be used in order to study the low density results of PRISM
theory 
analytically. For the mentioned case, $\nu=\frac 12$ in $d=3$, this was first
used in \cite{thread0,thread}, 
explicitly justified in \cite{threada}, and without
proof extended to exponents $\nu$ within the bounds $\frac 1d \le \nu < \frac
2d$  in \cite{fractal}. Here, general arguments 
on the solutions of \glto{2.1}{2.4} allow to find the low density
limits in the more general case $\frac 1d<\nu$ and $d\ge2$. 

The special low density limit, called ``thread PRISM'' model 
\cite{thread0,thread},  
studied in the following assumes that polymer solutions can be modeled as a
low density limit of the one--component PRISM equations for polymer
melts. Special solvent effects, are assumed to be taken into account via the
model for the intra--molecular structure, \glto{1.0a}{1.0c}. 

In general, the excluded volume constraint, \gl{2.1}, and the PY closure,
\gl{2.4}, lead to a discontinuity in the pair correlation function \gr 
and in the direct correlation function $c(r)$ 
at contact:
\beq{2.5}\begin{array}{lcll}
g_d & := & g(r\searrow R) & > 0\; ,\\
c(R-) & := & c(r \nearrow R) & \ne 0\; . 
\end{array}
\eeq
Note, that the actual values of $g_d$ and $c(R-)$
will dependent on details of the monomer
chemistry, as the segment size $R$ obviously is a microscopic length.
To connect both quantities it is useful to express
 \hq and $c_q$ as  one--dimensional Fourier transforms, 
\beq{2.5a}\begin{array}{lcl}
h_q & = &  \int dr\;  e^{i q r }\; j(r)\; ,\\
c_q & = &  \int dr\;  e^{i q r }\; i(r)\; ,\end{array}
\eeq
with the symmetric functions,
\beq{2.5b}\begin{array}{lcl}
j(r) & = & \Omega_{d-1} \int_{|r|}^\infty ds \; s\, (g(s)-1) \; 
(s^2-r^2)^{\frac{d-3}{2}}   \; ,\\
i(r) & = & \Theta(R-|r|)\; \Omega_{d-1} \int_{|r|}^R ds \; s\, c(s) \; 
(s^2-r^2)^{\frac{d-3}{2}}   \; , \end{array}
\eeq
where  $\Omega_d=\frac{2\pi^{d/2}}{\Gamma(d/2)}$
 denotes the surface of the $d$--dimensional unit sphere.
Because of \gls{2.1}{2.4}, $i(r)$ can be non--smooth
 for $|r| \le R$ only, while this can happen for $j(r)$ for $|r|\ge R$.
Thus, using the large wave vector limits, 
 $h_q\propto -g_d \cos{qR}/q^d$ for $q\to\infty$ and similarly for
$c_q$, and the large wave vector asymptote of \om, see \gl{1.0c}, \gl{2.2}
 connects the discontinuities of \gr and $c(r)$ at $r=R$, leading to:
\beq{2.6}
g_d =  - \, c(R-)\; .
\eeq
Moreover, one concludes that $c(r)$ is finite. Thus,  the scaling of the 
Fourier transform of the direct correlation function with global parameters (to
be defined below) also can be connected to the contact value:
\beq{2.7}
c_q = R^d c(R-) f_c(qR) =: - g_d R^d f_c(qR)\; ,
\eeq
where the regular function,\newline
$f_c(x)=\int_{y<1} d^dy\, e^{i {\bf xy}}\, c(R{\bf y})/c(R)$, 
has  a finite value at $x=0$. For the analytically known results of the
PRISM equations these property could be shown explicitly \cite{threada}, and
they can now be used to simplify the PRISM equation for low densities.

In order to extract the large distance solution of the PRISM equations, it
proves useful to shift the microscopic length scales to zero:
\beq{2.8} 
R\to0\, ,\quad \sigma\to0\qquad \mbox{with }\;  \sigma/R = \mbox{const.}\; .
\eeq
In order to evade the trivial limit of a non--interacting ideal gas, the length
scale of the intra--molecular correlations, to be denoted by $\xi_c$, which is
proportional to the molecular size, $\xi_c\propto R_g$, is kept finite by
increasing the index of polymerization $N$. 
\beq{2.9}
N\to\infty\qquad \mbox{so that }\; \xi_c \propto \sigma N^\nu  =
\mbox{const.}\; . 
\eeq
Also, in order to keep inter--molecular excluded volume active, the bare
segmental density 
is increased beyond bounds:
\beq{2.10}
\varrho\to\infty\qquad\mbox{so that }\; \varrho/\varrho_* = \mbox{const.}\; .
\eeq
In the thread PRISM equations, there enters a typical density, the dilute 
to semidilute crossover density $\varrho_*$, which is familiar from scaling
considerations 
\cite{degennes}. As will be shown below, $\varrho_*$, is defined differently
for 
scaling exponents $\nu$ below and above a value $\nu_c$, which denotes the
crossover to mean--field like  
behavior. 
\beq{2.11}
\varrho_* \propto \left\{
\begin{array}{lll}
 \frac{N}{\xi_c^d} & \mbox{for} & \nu < \nu_c = \frac 2d \; ,\\
\frac{1}{N \sigma^d} & \mbox{for}  & \nu > \nu_c  \; .
\end{array}\right.
\eeq
Equations  (\ref{2.8}) to (\ref{2.11}) specify the thread PRISM limit. Note,
that the divergent number density 
$\varrho_*$ actually corresponds to a vanishing polymer volume fraction,$\phi$,
 and thus (semi--) dilute 
polymer solutions are studied as claimed.
\beq{2.12}
\phi_* = \varrho_* \sigma^d  \propto \left\{
\begin{array}{lll}
1/N^{(\nu d-1)} & \mbox{for} & \nu < \nu_c \; ,\\
1/N & \mbox{for} & \nu > \nu_c\; ,
\end{array}\right.
\eeq
where $\phi={\cal O}(1)$ corresponds to polymer melts.

The solution of the non--linear PRISM integral equations in
the general case of course demands to find $f_c(x)$ from \gl{2.7} for all
$x$. However, a  solution to \glto{2.1}{2.4} depending on $f_c(0)$ can be
constructed on large distances, $r\gg R,\sigma$, or, equivalently in the thread
limit, for finite distances, $r>0$. 
This holds, because in the limit of $R\to0$, the excluded volume condition
\gl{2.1} affects a point (of measure zero) only, and thus \hq can be obtained
from the Fourier integral of \gr outside the core, $r>0$. From the inverse
transformation and \gls{2.2}{2.7}, the contact value follows in the general
case, where $\bf R$, denotes a vector of length $R+$: 
\beq{2.13}
g_d - 1  = - \frac{N A}{\varrho}\;
\int \frac{d^d q}{(2\pi)^d}\, e^{i{\bf qR}}\;  
\frac{\bar{f}_c(qR)\bar{\omega}_q^2}{1+A \bar{f}_c(qR) \bar{\omega}_q}\; .
\eeq
The normalized functions $\bar{\omega}_q=\omega_q/N$ and
$\bar{f}_c(x)=f_c(x)/f_c(0)$ have been introduced, and the
long wavelength interaction parameter A, which
abbreviates the zero wave vector limit of the direct correlation function:
\beq{2.14}
A = - N \varrho c_{q=0} =  N \varrho g_d R^d f_c(0) \; .
\eeq
For (semi--) dilute solutions, the limit \gl{2.8} simplifies the PRISM integral
equations, 
because the core condition, \gl{2.1}, becomes irrelevant, and \gl{2.13} 
assures that the connection \gl{2.6} is satisfied. The two further conditions,
\gls{2.9}{2.10}, assure that non--trivial solutions describing an interacting
polymer solution are obtained.
Because of \gl{2.9}, only the long ranged scaling form of the intra--molecular
 structure factor enters:
\beqa{2.15}
\bar{\omega}_q & = & \bar{\omega}(x=q\xi_c) \nonumber \\
 & = & \left\{\begin{array}{ll} 1+{\cal O}(x^2) & x\to0\; ,\\
                                1/x^{1/\nu} & x\to\infty \; .\end{array}\right.
\eeqa
And \gl{2.10} enforces the molecules to interact, $A\ne0$,
so that \gl{2.13} leads to a transcendental equation
determining $A$ (equivalently $g_d$), which is the only unknown parameter in
the thread inter--molecular structure factor: 
\beq{2.16}
\varrho h_q = - N A \frac{\bar{\omega}_q^2}{1+ A\bar{\omega}_q}\; .
\eeq

\section{Thread limit results} \label{sex3}

In the dilute to semidilute concentration region, the thread PRISM result for
\hq assumes the RPA like form, \gl{2.16}, where the interaction parameter 
$A$ needs to be found from \gl{2.13}. The total structure factor then also has 
 simple RPA--like form:
\beq{3.1}
S_q = N \frac{\bar{\omega}_q}{1+ A \bar{\omega}_q }\; ,
\eeq
where generally $\bar{\omega}_q$ differs from the Gaussian form and
the integrated inter--molecular interaction strength, $A$, in general
differs from the simple RPA approximation, $A\propto N\varrho R^d$.
Because of the large wave vector behavior of $\bar{\omega}$, \gl{2.15},
the limit $R\to0$ affects the integral in \gl{2.13} differently for 
$\nu<\nu_c=\frac 2d$ or $\nu>\nu_c$. In the first case, the integral converges
uniformly, 
and integration and limit can be interchanged, thus only $\bar{f}_c(0)=1$
enters. 
In the second case, the integral converges only because of the wave vector
dependence 
of $\bar{f}_c(qR)$, and leads to  RPA or mean--field behavior.

\subsection{Below the mean field crossover \label{sex3a}}

In the thread equation for the interaction parameter $A$, \gl{2.13}, 
the limit $R\to0$ can be performed trivially
for $\nu<\nu_c=\frac 2d$, and $A$ becomes a function of $\varrho/\varrho_*$
only, 
where the crossover density $\varrho_*=N\Omega_d/(2\pi\xi_c)^d$
enters.
\beq{3.2}
( 1 - g_d ) \frac{\varrho/\varrho_*}{A}  = 
\int_0^\infty dx \frac{x^{d-1} \bar{\omega}^2(x)}{1 + A \bar{\omega}(x)}\; .
\eeq
For size--mass scaling exponents $\nu$ corresponding to fractal dimensions,
$d_F=1/\nu$, equal to or exceeding  the spatial dimension, i.e. for $\nu<\frac
1d$,  the intra--molecular structure on long length scales determines the
thread equation (\ref{3.2}). There, screening of the intra--molecular excluded
volume has been neglected in the present approach, and the polymer segregation
effect predicted in the thread limit requires use of the self--consistent PRISM
approach \cite{fractal}. In order to avoid this complication, the exponent
$\nu$ will be restricted in the following, $\nu>\frac 1d$.
From the two limits of the integral in \gl{3.2},
  constant for $A\ll 1$ and $A^{-(2-\nu d)}$ for $A\gg 1$,  
the scaling form for the thread parameter can be determined.
\beqa{3.3}
A & = & (\varrho/\varrho_*) f_A(\varrho/\varrho_*) \; ,\qquad\mbox{where}
\nonumber \\
f_A(x) & \propto & \left\{\begin{array}{lll}
const. + {\cal O}(x) & \mbox{for} & x\to0\; ,\\
x^{\frac{2-\nu d}{\nu d-1}} & \mbox{for} & x\to\infty\; .
\end{array}\right. 
\eeqa
In \gl{3.3},  the contact value correction was already neglected, as it is
of higher order, as can be deduced from \gl{2.14}. Actually, a scaling law
follows 
from \gls{2.14}{3.3} for the contact value:
\beq{3.4}
g_d = c \frac{(\xi_c/R)^d}{N^2}\; f_A(\varrho/\varrho_*) \; ,
\eeq
where the identical scaling function $f_A$ from \gl{3.3} enters. The numerical
prefactor $c$ of course depends on the microscopic details of the polymer
model. 
Note that for $\nu<\nu_c$  the contact value vanishes like $N^{-(2-\nu d)}$ for
$N\to\infty$ in the dilute case.  
The scaling function $f_A$ also determines the density dependence of the mesh
size or density screening length, as for large reduced densities,
$\varrho\gg\varrho_*$,  the width of the total  
structure factor can be estimated from $S_q=(N/A)/(1+(q\xi_\rho)^{1/\nu})$,
for $q\xi_c\gg1$, which leads to:
\beq{3.5}
\xi_\rho = c' \sigma (\varrho\sigma^d)^{-\nu/(\nu d-1)}\; .
\eeq

For a given model of the intra--molecular structure factor, the thread
equation,  \gl{3.2} with $g_d=0$, allows to
determine $f_A$ straightforwardly. Figure \ref{fig1} shows the result for the
polymer model: $\omega_q=\frac 1N \sum_{\alpha,\beta=1}^N
\exp{-\frac{q^2\bar{\sigma}^2}{6}
 |\alpha-\beta|^{2\nu}}$, with the exponent given by
the Flory approximation $\nu=3/5$ corresponding to good polymer solutions 
\cite{degennes,descloi}. Numerical solutions
of the microscopic PRISM equations (\ref{2.1}) to (\ref{2.4}) for this model
and for not too large degrees of polymerization, $N$, still exhibit rather
large corrections to the thread asymptote.  This can be expected to be  model
dependent. From Fig. \ref{fig1} one notices that the connection of the contact
value to the small wave vector interaction parameter, \gl{2.14}, already holds
for values of $N$ where the asymptotic $f_A$, \gl{3.3}, is not yet
reached. Differences appear for larger packing fractions and signal
concentrated or melt like polymer packing.

The pair correlation function, $g(r)$, in the thread limit can be obtained for
finite segment distances from the Fourier transform of \gl{2.16}. For dilute
densities, $\varrho\ll\varrho_*$ and thus $A\ll1$, it differs from the ideal
gas  limit, $g(r)=1$, because of two molecule interactions. In the semidilute
concentration region, $\varrho\gg\varrho_*$ and $A\gg1$, the replacement
$A=(\xi_c/\xi_\rho)^{1/\nu}$ shows that $g(r)$ depends on the two length
scales, $\xi_c$ and $\xi_\rho$, independently.
On length scales large compared to the mesh size, $r\gg\xi_\rho$, the
inter--molecular structure factor exhibits the well known correlation hole
\cite{degennes,kcur1,kcur3,fractal}, which asymptotically for $\varrho\gg
\varrho_*$ exactly cancels off the long ranged
intra--molecular correlations:
\beq{3.5a}
h_q \to -\frac N\varrho\; \bar{\omega}(q\xi_c)\;,\fur q\ll 1/\xi_\rho\;;
\varrho\gg\varrho_* \; .
\eeq
This result is equivalent to $S_q\ll\omega_q$ for $q\xi_\rho\ll1$ in the
semidilute range \cite{degennes}. Note that the self--similar structure of the
molecule leads to the power law behavior $\varrho h_q \propto -(\sigma
q)^{-1/\nu}$ for $1/\xi_c\ll q\ll1/\xi_\rho$, which is equivalent to a power
law variation in the pair correlation function: $g(r)-1\propto (-1/\varrho)
(\sigma/r)^{d-1/\nu}$ for $\xi_\rho\ll r \ll\xi_c$ \cite{fractal}. For larger
distances, $r\gg\xi_c$, \gl{3.5a} describes how  $g(r)$  decays exponentially
to its random mixing value unity.

Within the mesh size, i. e. for distances around and smaller than the density
screening length, the inter--molecular correlations do not depend on the
molecular size and $\bar{\omega}$ in \gl{2.16} can be replaced by its large
$q\xi_c$ asymptote from \gl{2.15}. This leads to
\beq{3.5b}
h_q \to \frac{-\bar{c}\; 
\xi_\rho^d}{(q\xi_\rho)^{1/\nu} + (q\xi_\rho)^{2/\nu} }\; ,
\fur 1/\xi_c \ll q\; ; \varrho\gg \varrho_*\; ,
\eeq
where  the limiting behaviors of \hq for $q\xi_\rho $ large or small
compared to unity can be read of immediately and $\bar{c}=N/(\varrho A
\xi_\rho^d)$  approaches a number ($\bar{c}\to 6.26\ldots$ for $\nu=3/5$).
Note, that \gl{3.5} for the density screening length ensures a smooth crossover
of \gl{3.5b} to \gl{3.5a}  in the correlation hole region.
The variation of the pair correlation function, which describes the mesh
structure for $r\ll\xi_c$, can thus be obtained in closed form, if the neglect
of the cutoff of the correlation hole at $r\, \gtrapprox\,  \xi_c$ included in
\gl{3.5a}, is kept in mind. For $r\ll\xi_c$, $g(r)$ depends on $r/\xi_\rho$
only, with:
\beqa{3.5c}
g(r) & = & 1 - \bar{c} 
\int \frac{d^d y}{(2\pi)^d}\; e^{-i{\bf y r}/\xi_\rho} \;
\frac{1}{y^{1/\nu} + y^{2/\nu} } \nonumber \\
 & \to & \left\{ \begin{array}{lll}
(r/\xi_\rho)^{2/\nu-d} & r\ll \xi_\rho \; ; & r\gg \sigma,R\; , \\
1 - (\xi_\rho/r)^{d-1/\nu} & r\gg \xi_\rho \; ; & r\ll \xi_c\; , \\
\end{array}\right.
\eeqa
where $\bar c$ ensures $g(0)=0$ in agreement with \gl{3.2} and constant
prefactors of order unity have been suppressed in the final two lines.
For polymer chains in good solvents, the smooth increase, $g(r\ll\xi_\rho)
\propto (r/\xi_\rho)^{1/3}$, by accident agrees with the estimate from Ref. 
\cite{daoud}, $g(r\ll\xi_\rho) \propto
(r/\xi_\rho)^{(\gamma-1)/\nu}\approx(r/\xi_\rho)^{1/3}$ where
$\gamma$ is associated with the entropy of a single polymer chain
\cite{descloi}. 
The depth of the correlation hole displays an intriguing dependence on the
fractal and spatial dimensionalities. The probability to find a segment of
another polymer within the considered molecule decreases strongly if $1/\nu\to
d$. From \gl{3.5c} one estimates $g(r\approx \xi_c)-1\propto - 1/N^{\nu d-1}$,
which becomes a number of order unity in the case $\nu=1/d$. 
The smooth variation of $g(r)$ at short distances explains, why  the
scaling of the  correct  contact value $g_d$, \gl{3.4}, with macroscopic
variables can be estimated from the thread solution, \gl{3.5c}, by $g_d\propto
g(\sigma)$; its dependence on the ratio of the microscopic length scales,
$\sigma/R$, however cannot generally be recovered in this way \cite{threada}. 
Note that \glto{3.5}{3.5c} asymptotically apply for semidilute solutions,
$\varrho\gg\varrho_*$, whereas (\ref{2.16},\ref{3.1},\ref{3.3}) describe the
full dilute--to--semidilute crossover region.

\subsection{The mean field cases \label{sex3b}}

The condition \gl{2.13} for the contact value $g_d$, or equivalently, for
 the thread
parameter $A$ becomes independent of the microscopic interaction details only
for $\nu<\nu_c$. Above the crossover exponent, $\nu>\nu_c=\frac 2d$, the
integral over the effective 
potential as it enters \hq is determined by the local structure in
$f_c(qR)$. In the thread limit   
\gl{2.13} becomes a linear, density--independent equation for $g_d$
with solution: 
\beq{3.6}
g_d = 1 / [\, 1 + \int\frac{d^dq}{(2\pi/R)^d}\; e^{i{\bf qR}}\;
(q\sigma)^{-2/\nu} f_c(qR)\, ]\; . 
\eeq
Thus, a finite density independent contact value follows in the mean field like
cases $\nu>\nu_c$. 
Obviously, its exact value depends on the solution of the PRISM equations
considering all  
microscopic details and is beyond the reach of the thread PRISM approach.
The interaction parameter $A$ thus shows the density scaling as expected within
RPA.  
The reduced density $\varrho/\varrho_*$ appears, 
with $\varrho_*= 1/(N\sigma^d)$, and  $A$ becomes --- with a unknown 
numerical constant $\tilde c$, which however may depend on the ratio $\sigma/R$
of the microscopic  length scales.
\beq{3.7}
A = \tilde{c} \; \varrho/\varrho_*\; .
\eeq
In the semidilute density regime, the width of the total structure factor again
determines the density screening length,  $S_q=(N/A)/(1+(q\xi_\rho)^{1/\nu})$
for $q\xi_c\gg1$, with the result:
\beq{3.8}
\xi_\rho = \tilde{c}' \; \sigma (\varrho\sigma^d)^{-\nu}\; .
\eeq
Again, the numerical prefactor, $\tilde c'$, depends on the polymer model. The
result, \gl{3.5a}, discussed for the inter--molecular structure in 
the correlation hole region holds. The mesh structure factor,
$h(q\gg1/\xi_c)$, however, shows a different density scaling:
\beq{3.8b}
h_q \to \frac{-\hat{c}\, 
\sigma^d \,(\xi_\rho/\sigma)^{2/\nu}}{(q\xi_\rho)^{1/\nu} +
(q\xi_\rho)^{2/\nu} }\; , \fur 1/\xi_c \ll q\; ,
\eeq
which again, with \gl{3.8}, leads to a smooth crossover for intermediate
distances, $1/\xi_c\ll q \ll 1/\xi_\rho$, but to a small distance divergence of
the thread pair correlation function, $g(r)\propto -(\sigma/r)^{(d-2/\nu)}$ for
$r\ll\xi_\rho$. This results from the neglect of
the wave vector variation of the direct correlation function, \gl{2.7}, and
reinforces that the validity of the thread $g(r)$ is restricted  to
$r\gg\sigma$ for $\nu>\nu_c$, where $g(r)$ is still positive, as it must be by
definition.

\section{Discussion and comparison with other approaches \label{sex4}}

In this work, scaling limits appropriate for the dilute to semidilute
concentration regime of macromolecular solutions have been derived starting
from the microscopic 
PRISM integral equations. RPA like expressions for the total density
fluctuations, the structure factor \sq, \gl{3.1}, were given, where the density
scaling of the interaction parameter, $A$, was
deduced from the local excluded volume constraint, \gl{2.1}. The thread
interaction parameter is connected to the more familiar excluded volume
parameter, $v$, via $A=\frac{N_A M \varrho}{M_0^2} v$, where $N_A$ is Avogadro's
number, $M$ the molecular and $M_0$ the monomer weight. Effective density
dependent excluded volume parameters $v(\varrho)$ have often been used in
connection with RPA expressions \cite{daoud,ullman}, and \gl{3.3}
justifies this. 

The crossover of the single chain correlations to Gaussian large distance
behavior for $r\gg \xi_\rho$ due to intra--molecular excluded volume has been
neglected and would affect the model for \om, \gl{1.1}, and consequently the
thread results for large distances. Use of self--consistent PRISM
\cite{kcur3} to incorporate this would be required, but \gl{3.3} for the thread
parameter $A$ indicates that no change of its density scaling can be expected.

The crossover density $\varrho_*$, \gl{2.11}, arises from the full microscopic
PRISM equations as the relevant low density scale, and importantly, the
qualitatively different definitions in the mean field, $\nu>\nu_c=\frac 2d$,
and in the non--trivial cases, $\nu<\nu_c$, are recovered. Whereas for
$\nu<\nu_c$ the molecular crossover density, $c_*=\varrho_*/N$, is defined in
terms of the molecular size only, $c_*\propto1/R_g^d$ for $\nu>\nu_c$, in the
mean field cases also a microscopic length, the segmental hard core diameter
$R$,  enters, $c_*\propto1/(R_g^{2/\nu} R^{d-2/\nu})$ for $\nu>\nu_c$.
This indicates that inter--molecular steric interactions become important as
soon as the macromolecules fill space for $\nu<\nu_c$, whereas for the more
open molecules, $\nu>\nu_c$, much higher densities are required. 

For chain polymers, the upper critical dimension,  which
separates mean field and fluctuation dominated structures, agrees with
the renormalization group  results,
 $d_c=2/\nu_c=4$ \cite{degennes,descloi}. For rod polymers, $\nu=1>\nu_c$, the
 mean field like behavior underlies the successful Onsager theory
of the nematic transition  \cite{onsager}  and is generally argued to be true
\cite{shimada}. Note that in the studied PRISM theory orientational,
``nematic'', interactions are not treated correctly \cite{kcur3}, and thus a
nematic transition for rods is missed. Very recently, PRISM has been
generalized to treat oriented polymer fluids and the isotropic--nematic liquid
crystal transition \cite{pickett}. As the (isotropic) crossover density $c_*$
for rods is of the order of the nematic transition density \cite{onsager}, 
a suppression of nematic order is required to study the described isotropic
semidilute rod solutions experimentally. Networks of stiff semiflexible
molecules like actin may provide good systems \cite{sackmann}.

Of course, the full PRISM integral equations, which have been introduced to
study dense polymer systems with short ranged melt--like correlations
\cite{kcur1}, incorporate wave vector dependent corrections in
e. g. the effective interaction $c_q$, see \gl{2.3}, when $1/q$ approaches
local length scales.

From the compressibility, which is connected to the zero wave vector limit of
the total structure factor, the equation of state can be obtained, 
where $\Pi$ denotes the osmotic pressure \cite{hansen,kcur3}:
\beqa{4.1}
\frac{\Pi}{\varrho k_BT} & = & \frac 1N + \frac 1\varrho \int_0^\varrho
d\varrho'  \frac{A(\varrho')}{N} \nonumber\\
& \propto  & \left\{\begin{array}{lll}
\frac 1N\, ( 1  + {\cal O}(\varrho/\varrho_*) )
 & \varrho \ll \varrho_*\; , & \\
(\varrho\sigma^d)^{\frac{1}{\nu d-1}} & \varrho\gg\varrho_*\,,&\nu<\nu_c\;,\\ 
\varrho\sigma^d & \varrho\gg\varrho_*\,,&\nu>\nu_c\; , 
\end{array}\right.
\eeqa
when $A$ is given by \gl{3.3}.
The non--mean field behavior for $\nu<\nu_c$ for semidilute concentrations
\cite{fractal} agrees with the exact Des Cloizeaux result \cite{descloi},
and the  second virial coefficient, $1/(N\varrho_*)\propto R_g^d/N^2$,
recovers the picture of dilute polymer coils interacting like hard spheres of
radius $R_g$ \cite{aviksem}, but it does not vanish for dilute
$\Theta$--solvents, i. e. for $\nu=\frac 12$ in the present approach.
PRISM theory apparently correctly captures the leading
asymptotic behaviors, the free molecule limit $\Pi=(\varrho/N) k_BT$ for
$\varrho\ll\varrho_*$ and the $N$--independent  power law for
$\varrho\gg\varrho_*$, but the next to leading terms  are not described
correctly in general.

The structure of the polymer mesh in semidilute solutions, i. e. the
inter--molecular structure factor, \hq, on length scales of the order of the
density screening length, $\xi_\rho$, has not been conclusively discussed from
first principles calculations. The thread PRISM results for the non--mean field
like case of polymer chains in good solvents give explicit results,
\gls{3.5b}{3.5c}, which can be  compared to results from other approaches.

\subsection{Comparison with scaling considerations \label{sex4a}}

Detailed scaling law considerations of \hq in the limit $qR_g\gg1$ have been
presented in \cite{jannink} and can be directly compared with \gl{3.5b}. The
limit, $h_q\propto \xi_\rho^d$ for $1/R_g\ll q \ll 1/\xi_\rho$ strongly differs
from the  correlation hole behavior, $\varrho h_q=-(q\sigma)^{1/\nu}$ with
$\sigma$ the Kuhn'sche segment size, predicted by
the thread PRISM theory for this wave vector window. Physically, the
long ranged variation of \hq arises from the rearrangement of the polymer
mesh around a  molecule on distances up to the  molecule's
size. This adjustment compensates for the excess density due to the considered
molecule, leading to the small total density fluctuations expected 
for concentrated systems. The correlation hole has first been predicted
and discussed for polymer melts \cite{degennes}, but PRISM theory also
predicts it for semidilute solutions \cite{kcur1}, in agreement with scaling
considerations in \cite{daoud} but in disagreement with the
mentioned scaling picture presented in \cite{jannink}.  Intriguingly, PRISM
theory, \gl{3.5c}, recovers the tendency of macromolecules  to
segregate for $\nu=1/d$ as discussed for ideal chains in two dimensions
\cite{degennes}. 

In agreement with the thread PRISM result, a scaling law is postulated in
Ref. \cite{jannink} for the intermolecular structure factor inside the coil
radius, $h(q\gg 1/R_g) = \bar{h}(q\xi_\rho)$, which leads to the 
prediction $h_q\propto q^{-d}$ for $q\to\infty$ \cite{jannink}. As a  scaling
law can only hold for distances large compared to  the microscopic length
scales, $q\sigma\ll 1$, this result can be compared with the thread
scaling power law, $h_q\propto q^{-2/\nu}\xi_\rho^{d-2/\nu}$
  in \gl{3.5b}, and again differs.
Within thread PRISM the behavior of $h_q$ arises naturally as it matches
smoothly to the microscopic limit, $h_q\sim \frac{g_d}{q^d} \cos qR$ for $qR\gg
1$, because the contact value vanishes 
asymptotically, $g_d\propto (\sigma/\xi_\rho)^{2/\nu-d}$ 
for $\xi_\rho \gg \sigma$.
 This supports the
expectation in \cite{daoud}. Computer simulations could address this question
for polymer chain solutions as shown in section IV.C,
 where corrections to the low density scaling law,
\gl{3.4}, need to be considered which will arise due to finite packing
fractions. 

Accepting the existence of a scaling law for the contact value 
of macromolecules in solutions, then \gl{3.4} can be used to connect  the PRISM
results  to field theoretic calculations for two--polymer systems.

\subsection{Comparison with field theoretic calculations \label{sex4b}}

Field theoretic calculations which employ the mapping of the polymer problem
onto the $O(n\to0)$ magnetic model lead to numerous single chain results and
have recently been extended to provide information about the inter--molecular
structure factor on short length scales \cite{jannink,schaefer}.
In \cite{jannink}, the mentioned behavior $h_q\propto q^{-d}$ for $q\to\infty$
is recovered from the field theoretic calculation and used to support the
scaling picture discussed in the previous section. The implications for $g(r)$ 
can be compared with
another field theoretic calculation which studies the number of intersections
of two polymer chains \cite{schaefer}. Let $\Sigma_2(R_e)$ be the number of
intersections of two random (or self--avoiding) walks whose end--to--end
distance is $R_e$:
\beq{4.3}
\frac{\Sigma_2(R_e)}{V (4\pi\sigma^2 )^{d/2}} = 
\sum_{\alpha,\beta=1}^N \langle \delta({\bf r}^{(2)}_0-{\bf r}^{(1)}_0 - 
{\bf R}_e )\;
\delta({\bf r}^{(2)}_\alpha-{\bf r}^{(1)}_\beta ) \rangle\; .
\eeq
For intermediate distances $R_e$, $\sigma\ll R_e\ll R_g$, 
where the two polymers
overlap but local effects do not dominate $\Sigma_2$, the scaling
$\Sigma_2\propto
(\sigma/R_g)^{\omega_{12}(P)}$ is predicted, where the two molecule correction
to scaling exponent $\omega_{12}$ appears \cite{schaefer}. 
The contact value can now be obtained from $\Sigma_2$ by integrating over all
possible distances and (trivial) factors of normalization, as can be seen from
Eqs. (\ref{1.7},\ref{2.5},\ref{2.8}).
\beq{4.4}
g_d = \int d^dR_e\; \frac{\Sigma_2(R_e)}{(4\pi\sigma^2)^{d/2}N^2}\; .
\eeq
Using the results for two polymers from \cite{schaefer} to obtain the scaling
of the contact value in the dilute case, one finds:
\beq{4.5}
g_d^{\rm RG} 
\propto \frac{R_g^{d-\omega_{12}(P)}}{ N^2} \; ,\fur\; \varrho\to 0\; .
\eeq
The thread PRISM result, \gl{3.3}, differs from this in general, because in
PRISM theory the correction to scaling exponent is approximated to 
$\omega^{\rm thread}_{12}=0$. 
Its value in quadratic order in $\varepsilon=4-d$ is known, and the value
appropriate for polymer chains in good solvents turns out to
$\omega_{12}(G)=\frac 12 \varepsilon-\frac{19}{64}\varepsilon^2+\ldots\approx
0.40$ \cite{schaefer}, which can be compared to the PRISM and to the mean field
approximation, $\omega_{12}^{\rm RPA}=d-2/\nu$, which qualitatively
differs because it is negative. 
The thread approximation, $\omega_{12}=0$, is correct at and above the upper
critical dimension $d_c$. 

It appears difficult  to envisage simple forms of $g(r)$ which reconcile
 the prediction $h_q \propto q^{-d}$ for $q\to\infty$ \cite{jannink}
with the results for $\Sigma_2$ \cite{schaefer}, and the dilute limit of thread
PRISM theory qualitatively agrees with the later. 

\subsection{Comparison with Monte Carlo simulations \label{sex4c}}

Monte Carlo simulations are well suited to study  the inter--molecular
structure of polymer solutions but face the difficult challenge  to achieve a
clear separation of the three 
length scales, segment size $\sigma$ (or excluded volume size $R$), density
screening length $\xi_\rho$, and  molecular size $R_g$ (or molecular
correlation length $\xi_c$); see the discussion in \cite{yamakov}. Whereas in
\cite{yamakov}  in the range $1/R_g\ll q \ll 1/\xi_\rho$ a discrimination of
the 
two predictions, $h_q\propto$ const. from scaling considerations \cite{jannink}
and $h_q\propto q^{-x}$ with $x\approx 1/\nu$ (as follows from the PRISM
treatment of the correlation hole) 
appears possible and appears to support the latter, no clear 
conclusions about the exponent in the  asymptotic behavior, $h_q \to q^{-x}$
for $1/\xi_\rho\ll q \ll 1/R$, with $x=d=3$ (scaling picture),
 $x=4$ (RPA),  or $x=2/\nu\approx3.34$ (thread PRISM) were
possible.  Even recent large scale simulations of the bond fluctuation model
(BFM) \cite{carmesin,deutsch}
do not provide a conclusive test of the large $q$
dependence if $h_q$ is considered \cite{marcus}. Figure \ref{fig3}
shows data  from Ref. \cite{marcus}
 for semidilute solutions and rather large chain lengths,   $N=2048$,
 where $\xi_c=94$, $\xi_\rho= 14$ 
for $c/c_*=97.7$ and
$\xi_\rho=31.1$ for $c/c_*=25.6$, and the steric
 segment size is $R=2$ in units of the lattice constant of the BFM.
For a fit, the asymptotic thread PRISM prediction, \gl{3.5b}, is shifted by a
factor indicating that non--asymptotic corrections to $\bar c$ cannot be
neglected.

A clearer picture of the polymer mesh structure is provided by the pair
correlation function $g(r)$, which is the Fourier transform of $h_q$ and
asymptotically should follow \gl{3.5c} in the thread limit, $\sigma, R \ll
\xi_\rho$ and $\xi_\rho \ll \xi_c\propto R_g$. Figure \ref{fig2} shows the two
$g(r)$ for the above parameters, where $\xi_\rho$ is defined by collapsing the
simulation data onto the master curve at $g(r=\xi_\rho)=0.747$.
Note that this is an unfamiliar definition of $\xi_\rho$ which gives values
(theoretically) proportional to the standard ones.
 These values of
$\xi_\rho$ also produce the collapse of the $h_q$ onto a common curve shown in
the inset of Fig. \ref{fig2} and lead to a reasonable collapse of the pair
correlation functions onto a common master curve. Finite size
corrections enter from short distances because of the finite excluded volume
segment sizes, $R/\xi_\rho$. These corrections 
can also be understood as finite packing fraction corrections.
Large distance deviations from a common curve appear because of the finite
chain sizes, $\xi_c/\xi_\rho$. Nevertheless, the  short distance behavior of
$g(r)$ provides a sensitive test of the various predictions. 
The prediction $h_q\to 1/q^d$ would correspond to a logarithmic variation of
$g(r\to0)$, which appears to be ruled out by the data. Also the thread PRISM
prediction, $g(r\to0)\propto r^{1/3}$, appears incompatible with the data,
even if finite segment size corrections are approximated by a shift of the
$r$--origin. The RPA prediction for Gaussian polymers, $g(r\to0)-g_d\propto r$,
 of a linear increase in $r$, can describe the
data over small intervals (like $0.1 < g < 0.25$) 
but fails to account for the slight curvature of especially
the lower density curve. Moreover, the contact value of the RPA cannot  be
expected to vanish asymptotically if parameters appropriate for a fit to $h_q$
are  used.  An increase in the range of a power law fit to $g(r)$
at the lower density up to an interval  $0.1< g <  0.42$ is possible if the
following assumption about the pair correlation function for
 semidilute solutions is made: 
\beqa{final}
& g(r) \to  \bar{g}(r/\xi_\rho)\; \fur \xi_\rho\to\infty\;\; ; \;
\xi_\rho/\xi_c
\to 0 \;,& \\  
& \bar{g}(x\ll1)  \propto  x^{2/\nu-d+\omega_{12}(P)} \; .& \label{final2}
\eeqa
where for dilute cases the same power law with the replacement
$\xi_\rho\to\xi_c$ can be expected from scaling considerations. This power law
would match the scaling law for $g(r)$ for $r\to0$
smoothly to the calculated vanishing contact value $g_d$ from \gl{4.5}.
Note that such a matching is
predicted by PRISM. In \gl{final2} however, the exponent is corrected because
the correction to scaling exponent $\omega_{12}(P)$ is taken into account.
According to scaling arguments \cite{johner,schaefer}, 
there is  a term of the form of \gl{final2} present 
in the intra--molecular correlations also, although it is masked by
chain--end effects there.

The expected power law for good solutions, $g\propto r^{0.80}$ for $r\ll
\xi_\rho$ with $\nu=0.588$ and $\omega=0.40$ \cite{degennes,descloi,schaefer},
is compatible with the
simulation data, if a finite shift owing to a finite segment size is
anticipated. The power law $h_q\to 1/q^{2/\nu+\omega_{12}(P)}$ 
also is compatible with the data as shown in Fig. \ref{fig3}, but could less
be argued on data for $h_q$ only.

\section{Conclusions}

The thread PRISM results derived and discussed here justify earlier
phenomenological extensions of RPA like expressions. The density dependence of
the excluded volume parameter is derived from a microscopic incorporation of
inter--molecular  excluded volume and intra--molecular connectivity.
Various comparisons with rigorous field theoretic calculations show that
leading asymptotic predictions, even for non--mean field like situations, are
captured correctly in the PRISM integral approach. The correction to scaling
exponent, which appears in the molecular mass dependence of the contact value
of two polymers, provides a typical example where thread PRISM provides a much
better description than mean field theory but fails to describe all
non--trivial correlations. PRISM theory suggests useful concepts like the pair
correlation function $g(r)$ and predicts scalings laws which provide a
framework for the interpretation of data if the exponents are corrected.
Thread PRISM thus turns out rather useful for semidilute
solutions, where it explicitly describes the inter--molecular correlations of
the polymer mesh and results from more rigorous approaches are scarce. 
Moreover, as PRISM  theory is  successful for polymer melts, it provides
the unique possibility to approach polymer systems at all densities. 

\acknowledgments{We would like to thank Profs. K. S. Schweizer, G. Jannink, 
 K. Binder, A. Milchev and L. Sch\"afer and Dr. J. Baschnagel 
for many valuable discussions, and Dr. E. David for providing the programs to
solve the PRISM equations numerically.
This work was supported by the Deutsche
Forschungsgemeinschaft under Grants No. Fu 309/2-1 and Bi 317.}

\end{multicols}
\clearpage
\begin{figure}[h]
\caption[Figur eins]{\label{fig1}}
Scaling function $f_A(c/c_*)$ (bold solid curve)
of the thread interaction parameter
$A$ versus rescaled (molecular) 
polymer concentration in double logarithmic presentation for the Flory
exponent, $1/\nu=1.67$;
$c_*=1/(2\pi^2\xi_c^3)$ is the molecular overlap concentration. 
The curves labeled with degree of polymerization, $N$, are
results from microscopic PRISM calculations of the model described in the text,
where $\xi_c=0.28 N^\nu\bar{\sigma}$ is found with (\protect\ref{2.15}).
Full symbols give $f_A$ determined from $S_{q\to 0}$, and open symbols from the
contact value $g_d$ shifted by model dependent factors (0.088, 0.086, 0.085,
0.085 with increasing $N$). 
\end{figure}
\begin{figure}[h]
\caption[Figur eins]{\label{fig3}}
Inter--molecular structure factors $h_q$  from a Monte Carlo simulation of the
BFM for  polymers of length $N=2048$ ($\xi_c=94$ for $\nu=0.588$; all lengths
given in units of the lattice constant of the BFM). The data are taken 
from \protect\cite{marcus} and are shown scaled with $\xi_\rho$ 
determined from the corresponding  $g(r)$'s of Fig. \protect\ref{fig2}.
The $+$'s belong to the rescaled density $c/c_*=25.6$ and the $\times$'s to
$c/c_*=97.7$. 
The asymptotic thread PRISM result, Eq. (\protect\ref{3.5b}), shifted by a
correction factor 0.25 
is shown as solid line, whereas the dashed line indicates a power law,
$q^{-3.80}$ following from Eq. (\protect\ref{final2}). 
Bold solid lines mark where $q=1$.
\end{figure}
\begin{figure}[b]
\caption[Figur eins]{\label{fig2}}
Pair correlation functions $g(r)$ versus rescaled distance $r/\xi_\rho$ of the
BFM for the two
densities, $c/c_*=25.6$ (thin solid line, $\xi_\rho=31.1$) and
$c/c_*=97.7$ (thin dashed line, $\xi_\rho=14.0$) from  \protect\cite{marcus}.
The choice of $\xi_\rho$ collapses the curves at $g=0.747$. Circles mark the
contact values $g_d$. The asymptotic
thread PRISM prediction, Eqn. (\protect\ref{3.5b}) 
(thick solid line), and shifted according to a finite segment
size (thin dot--dashed curve)  are shown. A small $r$ asymptote,
$g=0.99((r-0.855 R)/\xi_\rho)^{0.80}$ according to Eq. (\protect\ref{final2})
is indicated by a long dashed curve, where
$R=2$ in units of the lattice constant is the excluded volume segment diameter
of the BFM, and a linear asymptote is given 
by a dotted line. The bold solid line denotes $\xi_c/\xi_\rho$
for the lower density. The inset enlarges the small--$r$ region showing the
shifted thread PRISM, the power law and the linear curves with the same line
types as in the main part.  
\end{figure}
\clearpage
\pagestyle{empty} 
\begin{figure}[h]
  \centerline{\hspace*{0.cm}\rotate[r]{\epsfysize=18.cm 
  \epsffile{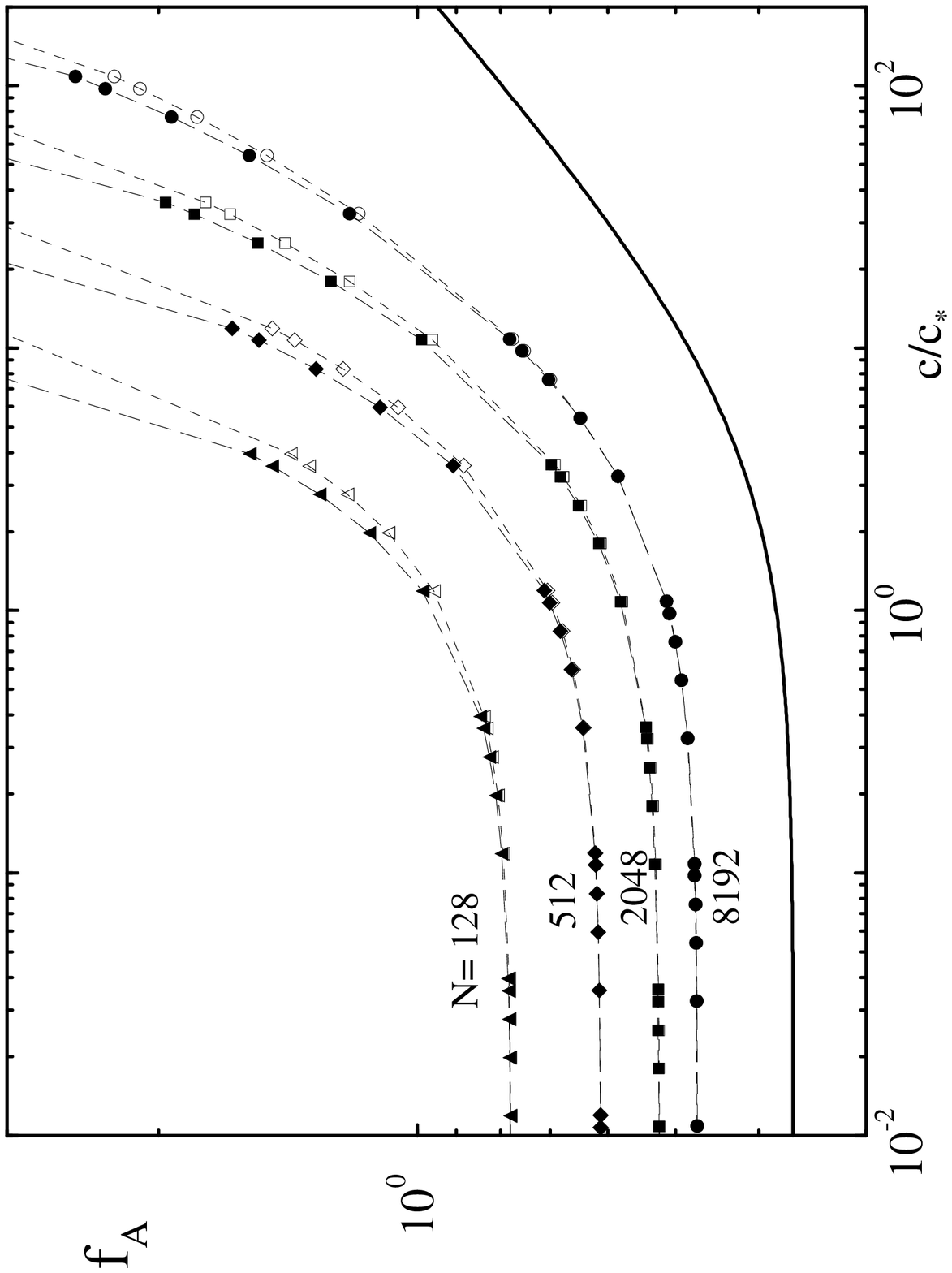}}}
  \vspace{2cm}
  \centerline{Fuchs and M\"uller, Inter--molecular ... Fig. 1}
\end{figure}
\clearpage
\pagestyle{empty} 
\begin{figure}[h]
  \centerline{\hspace*{0.cm}\rotate[r]{\epsfysize=18.cm 
  \epsffile{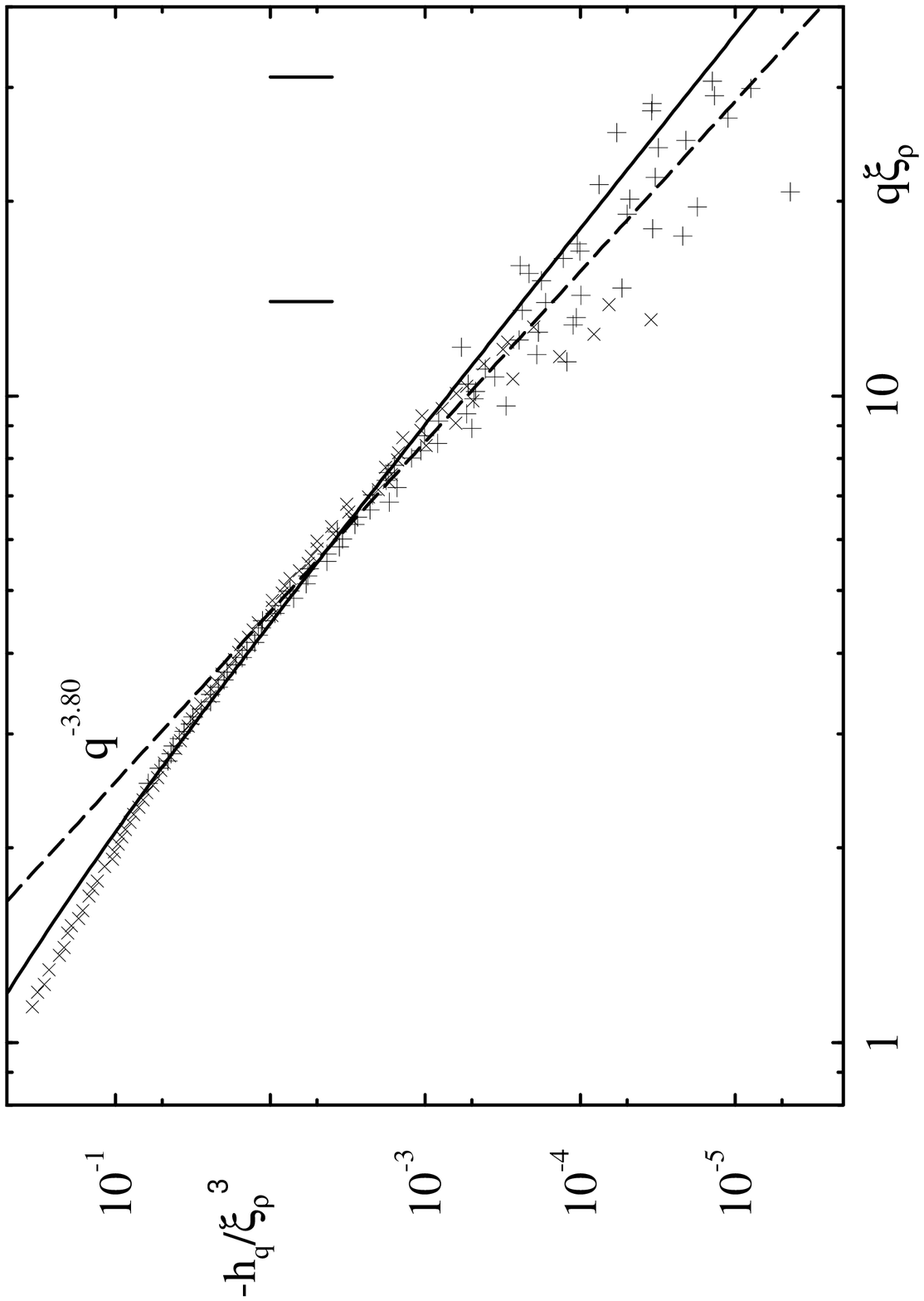}}}
  \vspace{2cm}
  \centerline{Fuchs and M\"uller, Inter--molecular ... Fig. 2}
\end{figure}
\clearpage
\pagestyle{empty} 
\begin{figure}[h]
  \centerline{\hspace*{0.cm}\rotate[r]{\epsfysize=18.cm 
  \epsffile{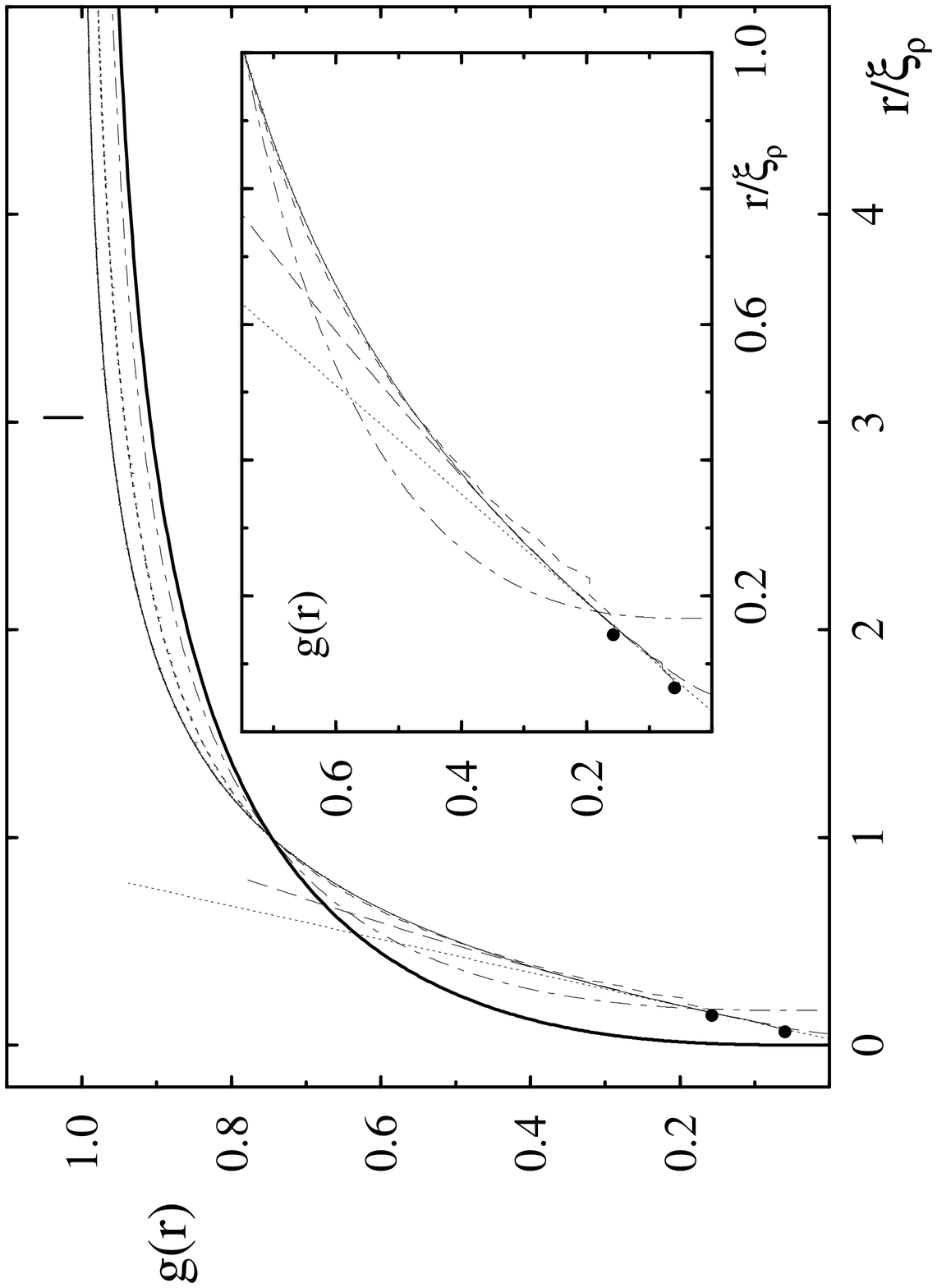}}}
  \vspace{2cm}
  \centerline{Fuchs and M\"uller, Inter--molecular ... Fig. 3}
\end{figure}
\end{document}